\documentclass[prl,a4paper,nofootinbib,
twocolumn,
]{revtex4}
\usepackage{graphicx}
\usepackage{amsfonts}
\usepackage{amssymb}
\usepackage{slashed} 
\usepackage{color}
\usepackage{hyperref}
\def\eq#1{{Eq.~(\ref{#1})}}
\def\eqs#1#2{{Eqs.~(\ref{#1})--(\ref{#2})}}
\def\fig#1{{fig.~(\ref{#1})}}
\def\sec#1{{sec.~(\ref{#1})}}
\def\fig#1{{fig.~(\ref{#1})}}
\def\tab#1{{tab.~(\ref{#1})}}
\def\tabs#1#2{{tabs.~(\ref{#1})--(\ref{#2})}}
\def\vev#1{\left\langle #1\right\rangle}
\def\abs#1{\left| #1\right|}
\def\mod#1{\abs{#1}}
\def\Im{\mbox{Im}\,}
\def\Re{\mbox{Re}\,}
\def\Tr{\mbox{Tr}\,}
\def\Str{\mbox{Str}\,}
\def\Det{\mbox{Det}\,}
\def\etc{{\it etc.}}
\def\etal{{\it et al.}}
\def\eg{{\it e.g.}}
\def\ie{{\it i.e.}}
\def\di{\mbox{d}}
\def\ltap{\ \raisebox{-.4ex}{\rlap{$\sim$}} \raisebox{.4ex}{$<$}\ }
\def\gtap{\ \raisebox{-.4ex}{\rlap{$\sim$}} \raisebox{.4ex}{$>$}\ }
\def\al{\alpha^{\prime}}
\definecolor{oucrimsonred}{rgb}{0.6, 0.0, 0.0}
\definecolor{persianblue}{rgb}{0.11, 0.22, 0.73}
\definecolor{forestgreen}{rgb}{0.13,0.35,0.13}
 \hypersetup{colorlinks, citecolor=oucrimsonred, linkcolor=persianblue, urlcolor=oucrimsonred}
\def\hhref#1{\href{http://arxiv.org/abs/#1}{#1}} 
\newcommand{\be}{\begin{equation}}
\newcommand{\ee}{\end{equation}}
\newcommand{\bea}{\begin{eqnarray}}
\newcommand{\eea}{\end{eqnarray}}
\newcommand{\nn}{\nonumber}
\newcommand{\spav}[1]{\parbox{1mmhe}{\vspace*{#1}}}
\newcommand{\spao}[1]{\mbox{\hspace{#1}}}
\newcommand\tw[2]{
\Bigg[\hspace{-1pt}\raisebox{1pt}
{$\begin{array}{c}
\displaystyle{#1} \\ \displaystyle{#2}
\end{array}$}
\hspace{-1pt}\Bigg]}
\newcommand\pertw[4]{
\Bigg[\hspace{-1pt}\raisebox{1pt}
{$\begin{array}{c}
\displaystyle{#1} \\ \displaystyle{#2}
\end{array}$}
\hspace{0pt}\Bigg|\hspace{0pt}\raisebox{1pt}
{$\begin{array}{c}
\displaystyle{#3} \\ \displaystyle{#4}
\end{array}$}
\hspace{-1pt}\Bigg]}
\newcommand{\U}{\scriptscriptstyle U}
\newcommand{\D}{\scriptscriptstyle D}
\newcommand{\s}{\scriptscriptstyle S}
\newcommand{\C}{\scriptscriptstyle C}
\newcommand{\N}{\scriptscriptstyle N}
\newcommand{\uu}{\scriptscriptstyle U\!U}
\newcommand{\dd}{\scriptscriptstyle D\!D}
\newcommand{\UR}{\scriptscriptstyle U_{\!R}}
\newcommand{\DR}{\scriptscriptstyle D_{\!R}}
\newcommand{\UL}{\scriptscriptstyle U_{\!L}}
\newcommand{\DL}{\scriptscriptstyle D_{\!L}}
\newcommand{\R}{\scriptscriptstyle R}
\newcommand{\LL}{\scriptscriptstyle L}
\newcommand{\Q}{\scriptscriptstyle Q}
\newcommand{\BR}{\mbox{BR\,}}

\begin{document}
\title[]{Dark-sector physics in the search for
 the rare decays  $K^+\rightarrow \pi^+  \nu \bar \nu$ and $K_L\rightarrow \pi^0  \nu \bar \nu$
}
\date{\today}
\author{M.\ Fabbrichesi$^{\dag}$}
\author{E.\ Gabrielli$^{\ddag\dag *}$}
\affiliation{$^{\dag}$INFN, Sezione di Trieste, Via  Valerio 2, 34127 Trieste, Italy }
\affiliation{$^{\ddag}$Physics Department, University of Trieste, Strada Costiera 11, 34151 Trieste}
\affiliation{$^{*}$NICPB, R\"avala 10, Tallinn 10143, Estonia }

\begin{abstract}
\noindent We compute the contribution of the decays  $K_L \rightarrow \pi^0 Q \bar Q$ and $K^+ \rightarrow \pi^+ Q \bar Q$, 
 where $Q$ is a dark fermion of the dark sector, to the measured widths for the rare  decays $K^+\rightarrow \pi^+  \nu \bar \nu$ and $K_L\rightarrow \pi^0  \nu \bar \nu$. The recent experimental  limit for $\Gamma (K^+ \rightarrow \pi^+ \nu \bar \nu)$ from {\sc NA62} sets a new and very strict bound on the dark-sector parameters. A branching ratio for $K_L \rightarrow \pi^0 Q \bar Q$ within the reach of the {\sc KOTO} sensitivity is possible. The Grossman-Nir bound is weakened by the asymmetric effect of the different kinematic cuts enforced by the NA62 and KOTO experiments. This last feature  holds true for all models where the decay into invisible states takes place through  a light or massless intermediate state.
\end{abstract}
\maketitle 
\section{Introduction}

The search for the rare decays $K^+\rightarrow \pi^+ \nu \bar \nu$ and $K_L\rightarrow \pi^0 \nu \bar  \nu$ is a most promising testing ground for physics beyond the standard model (SM) because their SM values are  ``short-distance'' dominated and can be predicted with great precision~\cite{Cirigliano:2011ny}. The contribution of many models beyond the SM to these decays has been studied (see, for example, the  review articles in~\cite{Buras:2004uu} and \cite{{Bryman:2011zz}}).

Among the models beyond the SM, those based on a dark sector containing  light dark fermions $Q$ (by definition singlet of the SM gauge groups and experimentally indistinguishable from the SM neutrinos)  are unique because they can introduce a contribution to these decays that is  a three-body decay (without the neutrinos in the final states) mediated by a massless vector boson. This  feature leads to the possibility of evading the Grossman-Nir (GN) bound~\cite{Grossman:1997sk} by means of a kinematical suppression which is asymmetric in the two decays.

 The idea that the width of the decay $K_L\rightarrow \pi^0 \nu \bar  \nu$ can exceed the value dictated by the GN bound  purely because of kinematical reasons is best illustrated by the following, rather extreme, case. 
 There exists a small region of the phase space where the decay $K^+ \rightarrow \pi^+ Q\bar Q$  vanishes  while the decay $K_L \rightarrow \pi^0 Q \bar Q$ remains open. This region is selected by taking values of $m_Q$   inside the interval
\be
m_{K^+} - m_{\pi^+} < 2 \, m_Q < m_{K_L} -  m_{\pi^0}\, . \label{quirky}
\ee
For $m_Q$ within the interval in \eq{quirky}, the $K^+$ cannot decay into a charged pion and the pair of dark fermions while the $K_L$, owing to its larger mass, can. 
 
 In a more  general (and perhaps more realistic) case, the width $\Gamma(K^+ \rightarrow \pi^+ Q\bar Q)$ can be suppressed by the events selection  in the experimental setting dedicated to its measurement  more than the width $\Gamma(K_L \rightarrow \pi^0 Q \bar Q)$ is by the  events selection applied in the corresponding experiment. Eventually, this asymmetry---which originates in the dependence of the signal events on the kinematical variables and their relationship to the experimental cuts---gives rise   to a $\Gamma(K_L \rightarrow \pi^0 Q \bar Q)$ larger than what  required to satisfy  the GN relationship.

In this paper, we analyze the two rare decays  $K^+\rightarrow \pi^+ Q \bar Q$ and $K_L\rightarrow \pi^0 Q \bar  Q$ in    a simplified model of the dark sector---which is inspired by dark sector scenarios in ~\cite{Gabrielli:2013jka,Gabrielli:2016vbb} and  contains new flavor changing neutral currents (FCNC) structures and CP violation independently of the SM.
Because of the asymmetric selection of the events outlined above, it is possible to bypass the GN bound and  obtain a  branching ratio $\BR (K_L \rightarrow \pi^0 Q\bar Q)$  compatible with all existing bounds on  FCNC physics and in the sensitivity range of the current experiments. Below a short summary of the experimental situation.

This year the upper bound in the result from {\sc BNL E949}~\cite{Artamonov:2008qb}
\be
\BR(K^+\rightarrow \pi^+ \nu \bar\nu) = 1.7\,^{+1.15}_{-1.05} \times 10^{-10}
\ee
has been (preliminarily) updated by {\sc CERN NA62}~\cite{NA62} to 
\be
\BR(K^+\rightarrow \pi^+  \nu \bar \nu) < 1.85 \times 10^{-10}\quad \mbox{90\% CL} \label{NA62}
\ee
which is now very close to the SM prediction which is~\cite{Buras:2015qea} 
\be
\BR(K^+\rightarrow \pi^+  \nu \bar \nu) = (7.81\,^{+0.80}_{-0.71}\pm 0.29 ) \times 10^{-11}\, ,
\ee
where the first error summarizes the parametric, the second the remaining theoretical uncertainties. 

 \begin{figure}[t!]
\begin{center}
\includegraphics[width=2.8in]{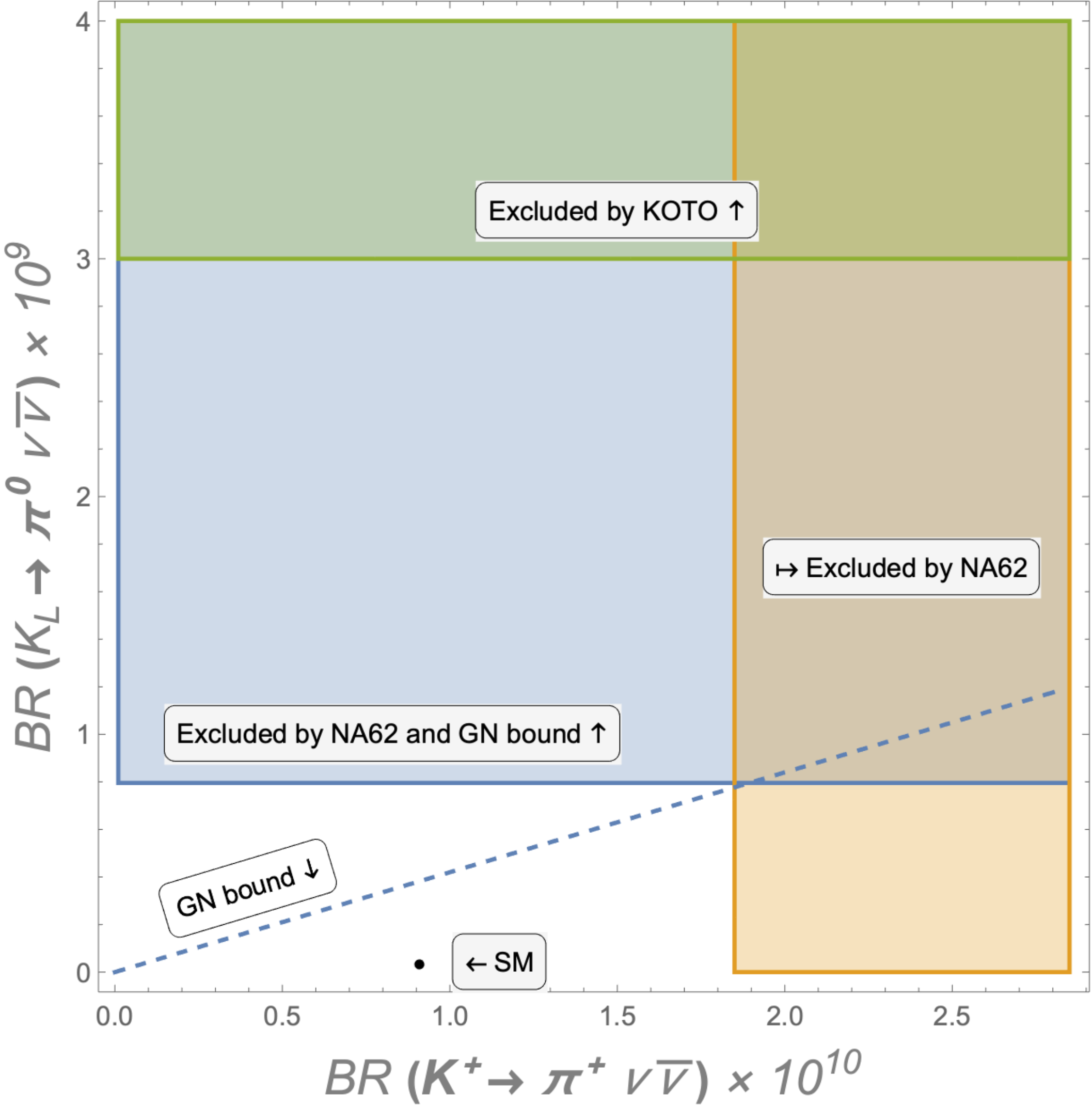}
\caption{\small 
\label{data} Summary of the experimental limits (90\% CL) on $K_L\rightarrow \pi^0  \nu \bar \nu$ ({\sc KOTO}) and  $K^+\rightarrow \pi^+  \nu \bar \nu$ ({\sc NA62}).
Also indicated are the GN bound and the SM predictions. The blue region is excluded  (assuming the validity of the GN bound). 
}
\end{center}
\end{figure}

 Meanwhile  the limit from the 2015 run at {\sc J-PARC KOTO}~\cite{Ahn:2018mvc} 
\be
\BR(K_L\rightarrow \pi^0  \nu \bar \nu) < 3.0 \times 10^{-9}\quad \mbox{90\% CL} \label{KOTOold}
\ee
is being  updated by data from the 2016-18 run with a single event sensibility (SES) of $6.9 \times 10^{-10}$~\cite{KOTO}. 
This SES spans a large range of values above the SM prediction, which is~\cite{Buras:2015qea} 
\be
\BR(K_L\rightarrow \pi^0  \nu \bar \nu) = (2.43\, ^{+0.40}_{-0.37} \pm 0.06 )\times 10^{-11}\, ,
\ee
where, as before,  the first error summarizes the parametric, the second the remaining theoretical uncertainties.

As shown in Fig.~\ref{data}, it is still possible that new physics dominates this channel and the current sensitivity of KOTO---falling as it does in the interval between the SM prediction and the exclusion limit in \eq{KOTOold}---could find it. Scenarios  giving rise to events in the KOTO SES range are discussed in~\cite{Kitahara:2019lws}.

Yet there is  a catch:
most of the range of the  SES  of KOTO and \eq{NA62}, when taken together,  violate the GN bound~\cite{Grossman:1997sk} 
\be
\BR(K_L\rightarrow \pi^0  \nu \bar \nu) \leq 4.3 \, \BR(K^+\rightarrow \pi^+ \nu \bar \nu)\, , \label{GN}
\ee
which is only based on isospin symmetry and the difference in the Kaon respective lifetimes. For this reason,
the very stringent new limit in \eq{NA62} on the charged Kaon decay seems to  imply a comparably stronger limit on new physics in the neutral Kaon channel, as depicted in Fig.~\ref{data} by the blue exclusion region. 

As anticipated,  this bound can be bypassed in the simplified dark-sector model by either the  vanishing of the $\BR(K^+\rightarrow \pi^+  Q \bar Q)$ when  the mass of the dark fermions is taken in the interval in \eq{quirky} or because of the different selections of the events in the kinematical regions explored by the two experiments. As discussed below, only the second possibility is fully consistent with the KOTO events.

The paper is organized as follows. In the next section we present the details of a model for the dark sector, including the most relevant constraints. In section 3 we give the predictions for the total decay width and branching ratio of $K_L\rightarrow \pi^0  Q \bar{Q}$, while in section 4 we analyze the impact of the experimental cuts selections on the branching ratios. Finally, in section 5 we present our conclusions.

\section{A  model of the dark sector}

Among the many models for the dark sector (see, for example, the review articles in~\cite{Raggi:2015yfk}), we use one made to resemble QED---that is, a  theory of charged fermions. It has the advantage of being simple. It contains  fermions $Q^{\U_i}$ and $Q^{\D_i}$, where  the index $i$ runs over generations like in the SM, and these dark fermions are charged only under a gauge group $U(1)_{\D}$---a proxy  for more general interactions---with different charges for the  $Q^{\U}$ and $Q^{\D}$ type. The  dark photon is massless and directly only couples to the dark sector~\cite{Dobrescu:2004wz} (in contrast with the case of massive dark photons).
We denote throughout with $\alpha_{\D}=e_{\D}^2/4 \pi$ the $U(1)_{\D}$ fine structure constant. 

There is no mixing between the ordinary and the dark photon because such a term in the Lagrangian can be rotated away~\cite{Holdom:1985ag,delAguila:1995rb} (again, in contrast with the case of the massive dark photon). The dark fermions carry an electric millicharge, the value of which is severely limited by existing constrains (see, for example, the relative discussion in~\cite{Beacham:2019nyx}). This millicharge and the dark photon coupling $e_{\D}$ are independent parameters and we consider the case in which the millicharge is negligible with respect to $e_{\D}$.

The dark model scenario we are using here is a simplified version of the models in ~\cite{Gabrielli:2013jka,Gabrielli:2016vbb}, where only the relevant interactions for the physical processes we are going to discuss are retained.
  The original proposal ~\cite{Gabrielli:2013jka} and  its extended version to left-right $SU(2)_L\times SU(2)_R$ gauge group ~\cite{Gabrielli:2016vbb}, has been mainly introduced to provide a natural solution to the flavor hierarchy puzzle of SM fermion masses. This model predicts the existence of dark fermions and messenger fields (with universal mass), the latter having the same quantum numbers of squarks and slepton of supersymmetric models. The additional requirement of an unbroken $U(1)_D$ gauge theory in the dark sector, under which both dark fermions and messenger fields are charged, has the benefit to maintain stable the dark fermions (provided the messenger sector is heavier) thus promoting them to potential dark matter candidates.

The dark fermions couple to  the SM fermions by means of a Yukawa-like interactions. The Lagrangian contains terms coupling SM fermions of different generations  with the dark fermions. In general the interaction is not diagonal in flavor and, for the SM  $s$ and $d$  quarks relevant for Kaon physics, is given by
\be
{\cal L}  \supset 
g_R  \rho_R^{sd} S^{\dag}_R \bar{Q}^{d}_L s_R   +  g_L  \rho_L^{sd} S^{\dag}_L \bar{Q}^{s}_R d_L \label{Lsd}
 + H.c.  
\ee
 In \eq{Lsd}, the fields $S_{L}$ and $S_{R}$ are  heavy messenger scalar particles, doublets and singlets of the SM $SU_L(2)$ gauge group respectively as well as  $SU(3)$ color triplets (color indices are implicit in \eq{Lsd}). 
The  symmetric matrices $\rho^{sd}_{L,R}$ are the result of  the diagonalization of the mass eigenstates of both the  SM and dark fermions; they provide the generation mixing (and the CP-violation phases) necessary to have the messengers play a role in flavor physics.  
The messenger  fields are heavier than the dark fermions and charged under the $U(1)_{\D}$ gauge interaction, carrying the same charges as the dark fermions. 

 In order to fix the notation,  we report below also the Lagrangian for the flavor diagonal interaction 
\be
{\cal L}  \supset 
g_R  \rho_R^{ss} S^{\dag}_R \bar{Q}^{s}_L s_R   +  g_L  \rho_L^{dd} S^{\dag}_L \bar{Q}^{d}_R d_L \label{Lsd}
 + H.c.  \, .
\ee
The minimal flavor violation hypothesis requires the diagonal couplings $\rho$ to be $\rho_{L,R}^{ss},~\rho_{L,R}^{dd}\simeq 1 $ \cite{Gabrielli:2016cut}.

In order to generate chirality-changing processes, we also need in the Lagrangian the mixing terms
 
\be
{\cal L} \, \, \supset \, \, \lambda_S S_0 \left(  S_L  S_R^{\dag} \tilde H^\dag +  S_L^{\dag} S_R H \right) \, , \label{mix} 
\ee
where $H$ is the SM Higgs boson, $\tilde{H}=i\sigma_2 H^\star$, and $S_0$ a scalar singlet. The Lagrangian in \eq{mix} gives rise to  the mixing after the scalars $S_0$ and $H$ take a vacuum expectation value (VEV), respectively, $\mu_S$  and $v$---the electroweak VEV. 
After diagonalization, the messenger fields $S_\pm$ couple both to left- and right-handed SM fermions with strength $g_L/\sqrt{2}$  and $g_R/\sqrt{2}$, respectively. We can assume that the size of this mixing---proportional to the product  $\mu_s v$ of the VEVs---is large and of the same order of the masses of the  scalars. 

This model (see \cite{Gabrielli:2016vbb} for more details)  has been used to discuss processes with the emission of dark photons in Higgs physics~\cite{Biswas:2016jsh}, flavor changing neutral currents~\cite{Gabrielli:2016cut}, kaon~\cite{Fabbrichesi:2017vma,Barducci:2018rlx} and $Z$ boson~\cite{Fabbrichesi:2017zsc}  decays.

\subsection{Coupling SM fermions to the dark photon}
\begin{figure}[t!]
\begin{center}
\includegraphics[width=3.1in]{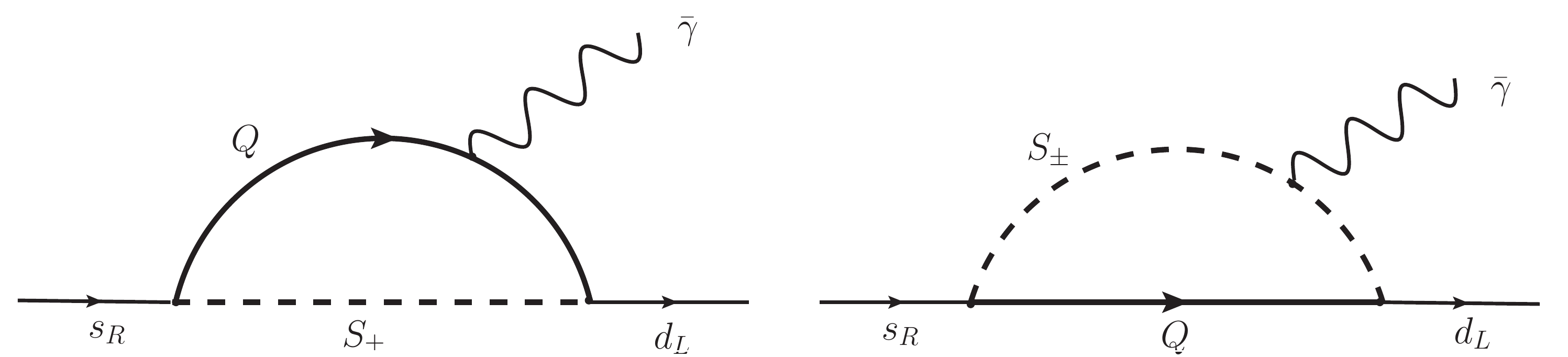}
\caption{\small Vertex diagrams for the generation of the dipole operators in the  model of the dark sector. 
\label{vertex} }
\end{center} 
\end{figure}

  SM fermions couple to the dark photon only via non-renormalizable interactions~\cite{Dobrescu:2004wz} induced by loops of  dark-sector  particles. 
The corresponding effective Lagrangian relevant for the rare decays of the Kaons is equal to 
\be
{\cal L}= \frac{e_{\D}}{2\Lambda} \bar{s} \, \sigma_{\mu\nu} \left( {\cal D}_M + i \gamma_5 {\cal D}_E \right) d \,  B^{\mu\nu} + H.c. \, ,\label{dipole}
\ee
where $B^{\mu\nu}$ is the strength of the dark photon field, $\Lambda$ the  effective scale of the dark sector, which is the same order of magnitude as the scalar masses $m_S$. The magnetic- and electric-dipole are given by 
\be
{\cal D}_M =\frac{ \rho_{sd} \rho_{dd}^*}{2} \, \Re \frac{g_L g_R}{(4 \pi)^2} \; \; \mbox{and}  
\;\;  {\cal D}_E = \frac{ \rho_{sd} \rho_{dd}^*}{2} \, \Im \frac{g_L g_R}{(4 \pi)^2}   \, , \label{ds}
\ee
respectively. For simplicity we take $g_L=g_R$ real and $ {\cal D}_E =0$. A CP-violating phase  comes from the mixing parameters:
\be
 \rho_{sd} \rho_{dd}^* - \rho_{sd}^* \rho_{dd} = 2\,  i \sin \delta_{\mbox{\tiny CP}} \, . \label{phase}
\ee

\subsection{Constraints on the parameters of the model}

The size of the coupling $\alpha_{\D}$ is  constrained by galaxy dynamics and cosmology (see \cite{Ackerman:mha,Feng:2009mn,Agrawal:2016quu}) if dark matter is among the fermions charged under $U(1)_{\D}$. This  limit depends on the mass of the dark matter.  The coupling $\alpha_{\D}$  can be as large as 0.1 for a mass around 10 TeV, while values around $\alpha_{\D} =0.001$ (like those we shall use) require a mass around 100 GeV.

Anyway, the light dark fermions $Q$ into which the dark photon decays in 
$K_L \rightarrow \pi^0 Q \bar Q$ and $K^+ \rightarrow \pi^+ Q \bar Q$
are not dark matter because they  have  annihilated before the current epoch   into dark photons $\bar \gamma$ with a thermal averaged cross section given by
\be
\langle \sigma_{Q \bar Q \rightarrow \bar \gamma \bar \gamma} v \rangle = \frac{2 \pi \alpha_{\D}^2}{m_\chi^2}  \, ,\label{thermal-x-section1}
\ee
where $v$ is the relative velocity of the annihilated pair.
For a  strength $\alpha_{\D}$ in the range we shall use (namely, between 0.0003 and 0.004, see Fig.~\ref{br0} below), all dark fermions with masses  of order 100 MeV have a large  cross section and their relic density
\be
\Omega_\chi\, h^2 \approx \frac{2.5 \times 10^{-10} \; \mbox{GeV}^{-2}}{ \langle \sigma_{Q \bar Q \rightarrow \bar \gamma \bar \gamma} v \rangle}
\ee
is below $10^{-6}$ and therefore negligible.

The scale $\Lambda$ is  constrained by astrophysical and cosmological data~\cite{Dobrescu:2004wz,Giannotti:2015kwo}. These limits only refer to the flavor conserving interactions---mostly electrons in the case of stellar cooling, muons and $s$-quark in primordial nucleosynthesis and light quarks in the 1987A supernova explosion---and we assume here that they are not relevant because do not apply  in our flavor-changing process case. The only relevant limit is the one from Kaon mixing that we include in our analysis by means of \eq{limit} below.

There are no bounds on the masses $m_Q$ of the dark fermions because of their very weak interaction with the SM states. There may be a question about a light  mass for the dark fermion because of the impact of the dark sector on the cosmic microwave background. This point needs to be investigated further~\cite{io}.

Laboratory limits apply to the mass of the messenger scalar states $m_S$ of the model, which  is of the same order as  $\Lambda$. 
The messenger states have the same quantum numbers and spin of the supersymmetric squarks.
At the LHC they are copiously produced in pairs through QCD interactions and decay at tree level into a quark and a dark fermion. The final state arising from their decay is thus the same as the one obtained from the $\tilde q \to q \chi^0_1$ process.
Therefore limits on the messenger masses can be obtained by reinterpreting supersymmetric searches on first and second generation squarks decaying into a light jet and a massless neutralino~\cite{Aaboud:2017vwy}, assuming that the gluino is decoupled. 
In particular we have used the upper limits on the cross section for various squark masses of~\cite{Aaboud:2017vwy} that the ATLAS collaboration provided on {\tt HEPData}.  These limits have been used to compute the bounds as a function of the messenger mass  using next-to-leading order QCD cross section for squark pair production from the LHC Higgs Cross Section Working Group~\footnote{Available at the web-page~\url{https://twiki.cern.ch/twiki/bin/view/LHCPhysics/SUSYCrossSections}.}.

We take into account the contributions to the total event yield given only by right-handed (degenerate) messengers associated to the first generation of SM quarks, with the others set to a higher mass and thus with a negligible cross section. This corresponds to have only 2 light degrees of freedom, which are analogous to $\tilde u_1$ and $\tilde d_1$ in supersymmetry. With  this assumption we obtain a lower bound on their masses of 940 GeV, limit that increases up to 1.5 TeV by assuming that messengers of both chiralities associated to the first and second generation of SM quarks are degenerate in mass.

These limits on $\Lambda$ of the order of 1 TeV are much weaker than those obtained in the next section from the Kaon mass difference.


\subsection{Constraint from the Kaon mass difference}

A direct constraint on the parameters of the model arises because the same term driving   the meson decay also enters the box diagram  that gives  rise to the mass difference of the neutral meson. This quantity is given by  
  \bea
 \Delta m_{K^0} \!=\! \left[  \frac{g_L^{4} (\rho_{sd}^L)^2 \rho_{dd}^L \rho_{ss}^L +  g_R^{4}(\rho_{sd}^R)^2 \rho_{ss}^R \rho_{dd}^R }{\Lambda^2} \right] \!\!\frac{ f_{K^0}^2 m_{K^0}}{192 \pi^2} \, \label{DeltaM} 
 \eea
 where we have identified $m_S=\Lambda$ and used  the leading vacuum insertion approximation ($B_{K^0}=1$) to estimate the matrix element
 \bea
 \langle K^0 | (\bar{s}_L\gamma^\mu d_L) \,  (\bar{s}_L\gamma_\mu d_L) |\bar K^0 \rangle = \frac{1}{3} m_{K^0} f_{K^0}^2 B_{K^0} \eta_{\mbox{\tiny QCD}}
 \eea
and a  similar one for right-handed fields,
where $s_{L},d_{L}$ represent the corresponding quark fields with left-handed chirality.
Since we are just after an order of magnitude estimate,  we neglect the running (and contributions from mixing) of the Wilson coefficient $\eta_{\mbox{\tiny QCD}}$ of the 4-fermion operator. Given the long-distance uncertainties, to satisfy the experimental bound on the mass difference, we  only impose that the new contribution does not exceed the measured value. In order to simplify the analysis, in the expression of Eq.(\ref{DeltaM}) we have neglected the CP violating contributions, and, as already mentioned, assumed all couplings to be real and $g_L=g_R$. Moreover, to directly constrain the magnetic dipole interactions, we approximate the diagonal couplings $\rho_{dd}=\rho_{ss}=1$ in Eq.(\ref{DeltaM}), which is also in agreement with the minimal flavor violation hypothesis. 
  
  The comparison requires the introduction of the full effective Lagrangian~\cite{Gabbiani:1996hi} inclusive of the new operators induced by the dark sector. By using the results in \cite{Ciuchini:1998ix}, we obtain~\cite{Fabbrichesi:2017vma,Barducci:2018rlx}
  
  \be
  \frac{\left|{\cal D}_M\right|^2}{\Lambda^2 } \leq   \frac{3}{32 \pi^2} \frac{\Delta m_{K^0}^{\rm exp}}{f_K^2m_{K^0}}= 2.6 \times 10^{-21} \, {\rm MeV}^{-2} \, ,\label{limit}
  \ee
  with $f_K=$159.8 MeV and $\Delta m_{K^0}^{\rm exp}= 3.52 \times 10^{-18}$ MeV~\cite{PDG}.


\section{The decay width}

 This process is experimentally seen as  two photons (from the decay of the pion) plus the missing energy and momentum carried away by the neutrinos. In the presence of the dark sector, the same signature would be provided by $K\rightarrow \pi \bar\gamma$, where $\bar \gamma$ is a dark photon,  but this decay is forbidden by the conservation of the angular momentum when the dark photon is massless. This means that the decay we are interested in can only proceed if the dark photon is off shell and decays into a pair of dark fermions.  

 This signature could proceed also via  box diagrams at 1-loop, where in the internal states are running messengers and dark-fermions fields.
However, the box diagrams are suppressed---doubly,  by an extra mass factor $O(m_K / \Lambda)$ and  an additional factor $O( g_L^2 g_R^2 /(4\pi)^4/\alpha_D )$ with respect to the diagram with an off-shell dark-photon, thus they are subleading and we neglect them in our analysis.

Assigning the momenta as
\bea
K_L(p_K) \rightarrow \pi^0(p_\pi) Q(q_1) \bar Q(q_2)\, ,
\nonumber
\eea
we find
\begin{widetext} 
\bea
\frac{d \Gamma (K_L \rightarrow \pi^0 Q \bar Q)}{d z_1 d z_2}   & = & \frac{2 \alpha_D^2}{ \pi}   \frac{\left| {\cal D}_M \right|^2}{\Lambda^2}   \frac{ m_K \; |f^{K\pi}_T (z_1, z_2)|^2 
  \Omega_C (z_1,z_2) \sin^2 \delta_{\mbox{\tiny CP}}}{(1+r_\pi)^2}
\Big[ r_{\pi}^4 + 4 z_1 z_2 + r_{\pi}^2 (2 z_1 + 2 z_2 -1) \Big] \label{Gamma} 
 \, ,
\eea
\end{widetext}
where $r_{\pi}=m_{\pi}/m_K$, $z_1=q_1.p_\pi/m_K^2$ , $z_2=q_2.p_\pi/m_K^2$ and $\sin \delta_{\mbox{\tiny CP}}$ is defined in \eq{phase} and comes from the CP-violation  in the dark sector.  We have found {\sc Package-X}~\cite{Patel:2015tea} useful in checking \eq{Gamma}.

The Sommerfeld-Fermi  factor~\cite{sommerfeld} is given by
\be
\Omega_C (z_1,z_2) = \frac{\xi(z_1,z_2)}{e^{\xi(z_1,z_2)}-1}\, ,
\ee
with  
\be
 \xi (z_1,z_2)= -\frac{2 \pi \alpha_D }{\sqrt{1 - 4 m_Q^4/(q^2 - 2 m_Q^2)^2 }}\, ,
 \ee
  and $q^2 = m_K^2 - m_\pi^2 - 2 m_K^2 (z_1+z_2)$, arises from the (dark) attractive Coulomb interaction of the (dark) final states. This factor can be numerically important and partially compensates the kinematical suppression due to the smallness of the available phase space when $m_Q$ is sufficiently large; it is characteristic of having a dark sector with QED-like interactions.

In \eq{Gamma} we have taken for the hadron matrix element 
\be 
\langle \pi^0 | \bar s\,\sigma^{\mu\nu} d\, | K^0 \rangle = (p^\mu_\pi p^\nu_K - p^\nu_\pi p^\mu_K) \frac{\sqrt{2} f_T^{K\pi} (q^2)}{m_\pi + m_K} \, ,
\ee
where the tensor form factor is given  by 
\be
f_T^{K\pi} (q^2) = \frac{f_T^{K\pi} (0)}{1 - s_T^{K\pi} q^2} \, ,
\ee
with  $q^2$ as before and $ f_T^{K\pi} (0)=0.417(15)$ and $ s_T^{K\pi}=1.10(14)\; \mbox{GeV}^{-1}$ on the lattice~\cite{Baum:2011rm}.

The phase-space integration  is between 
\bea
z_1^{\rm min \atop \rm max}&=&\frac{(m_{12}^2)^{\rm min \atop \rm max}-m_{\pi}^2-m_Q^2}{2m_K^2} \nn \\
z_2^{\rm min \atop \rm max}&=&\frac{(m_{23}^2)^{\rm min \atop \rm max}-m_{\pi}^2-m_Q^2}{2m_K^2} \, , \nn
\eea
where
$$
(m_{12}^2)^{\rm min \atop \rm max}={(m_Q+m_{\pi})^2 \atop (m_K-m_Q)^2}
$$
and
$$
(m_{23}^2)^{\rm min \atop \rm max}=(E_2+E_3)^2-\left(\sqrt{E_2^2-m_{\pi}^2}\pm\sqrt{E_3^2-m_Q^2}\right)^2
$$
for
$$
E_2=\frac{\left(m_{12}^2-m_Q^2+m_{\pi}^2\right)}{2\sqrt{m^2_{12}}} \, ,\quad
E_3=\frac{\left(m_K^2-m_{12}^2-m_Q^2\right)}{2\sqrt{m^2_{12}}} 
$$
with $m_{12}^2=2m_K^2z_1+m_{\pi}^2+m_Q^2$.

The result for $\Gamma (K^+ \rightarrow \pi^+ Q \bar Q)$ is the same as that in \eq{Gamma} but  for the absence of the CP-violating $\sin^2 \delta_{\mbox{\tiny CP}}$ and for a factor 0.954 coming from the isospin rotation and the difference in the masses. The two widths together satisfy the GN relationship in \eq{GN} once the different lifetimes of the $K^+$ and the $K_L$ are taken into account in the BRs.

\subsection{$\BR(K_L\rightarrow \pi^0 Q \bar Q )$ without experimental cuts}

We take $m_{K_L}=497.61$, $m_{\pi^0}=134.98$ MeV~\cite{PDG} and span the possible values within the window in \eq{quirky} $178<m_Q< 181$ MeV assuming maximal CP violation ($\sin \delta_{\mbox{\tiny CP}}=1$). We vary the dark-photon coupling constant:  $0.05<\alpha_D<0.15$. After
enforcing the limit in \eq{limit}, we obtain that
the integration of \eq{Gamma} over the phase space leads to 
\be
3.9 \times 10^{-12}  < \BR(K_L\rightarrow \pi^0 Q \bar Q ) <  3.7 \times 10^{-8} \, ,   \label{result}
\ee
a range that covers the entire region  from below the SM prediction to above the KOTO SES region.

The result in \eq{result} only depends in a significative manner on 
\begin{itemize}
\item \underline{the choice of $m_Q$ and $\alpha_D$}. By taking $m_Q$ closer to the upper end of the window in \eq{quirky} we close the phase space and in the end, Sommerfeld-Fermi enhancement notwithstanding, the width goes to zero. Notice that the window in \eq{quirky} can be (slightly) enlarged by having the $\Gamma(K^+\rightarrow \pi^+ Q \bar Q )$  not closed but only suppressed by the kinematics below the experimental limit in \eq{NA62} (and still above the SM prediction). 

\item  \underline{$\sin \delta_{\mbox{\tiny CP}}$}. The whole decay width is proportional to the size of CP violation. Its size can be modulated by taking $\sin \delta_{\mbox{\tiny CP}}$ smaller than one.
\end{itemize}

 Notice that the new limit in \eq{NA62} would imply a strong bound on the dark-sector parameters if the channel were to be open and not kinematically restricted. We use this constraint in the next section.

\subsection{The transverse momentum of the pion}

The particular kinematic window in \eq{quirky} constrains the possible transverse momenta $p_T^{\pi^0}$ of 
the $\pi^0$ and we have
$$
p_T^{\pi^0} < \frac{\sqrt{[m_K^2 - (2 m_Q-m_{\pi^0})^2][m_K^2 - (2 m_Q+m_{\pi^0})^2]}}{2 m_K}
$$
which gives $p_T^{\pi^0} <36$ MeV---for the most favorable case of taking $m_Q=178$ MeV.  This value can be increased  to around 60 MeV if we allow $m_Q$ to drop below the threshold for the $K^+$ decay while still suppressing the width of this channel by the smallness of the phase space. 

The signal region of KOTO  cuts off pions with momenta smaller than 130 MeV to reduce the background from $K_L\rightarrow \pi^+\pi^-\pi^0$~\cite{Ahn:2018mvc}. 
 It is a prediction of the scenario with the choice in \eq{quirky} that the pions  have small transverse momentum and are, therefore,  in a  kinematical region  excluded by the KOTO experiment.

\section{Enter the experimental cuts}

 \begin{figure*}[ht!]
\begin{center}
\includegraphics[width=2.2in]{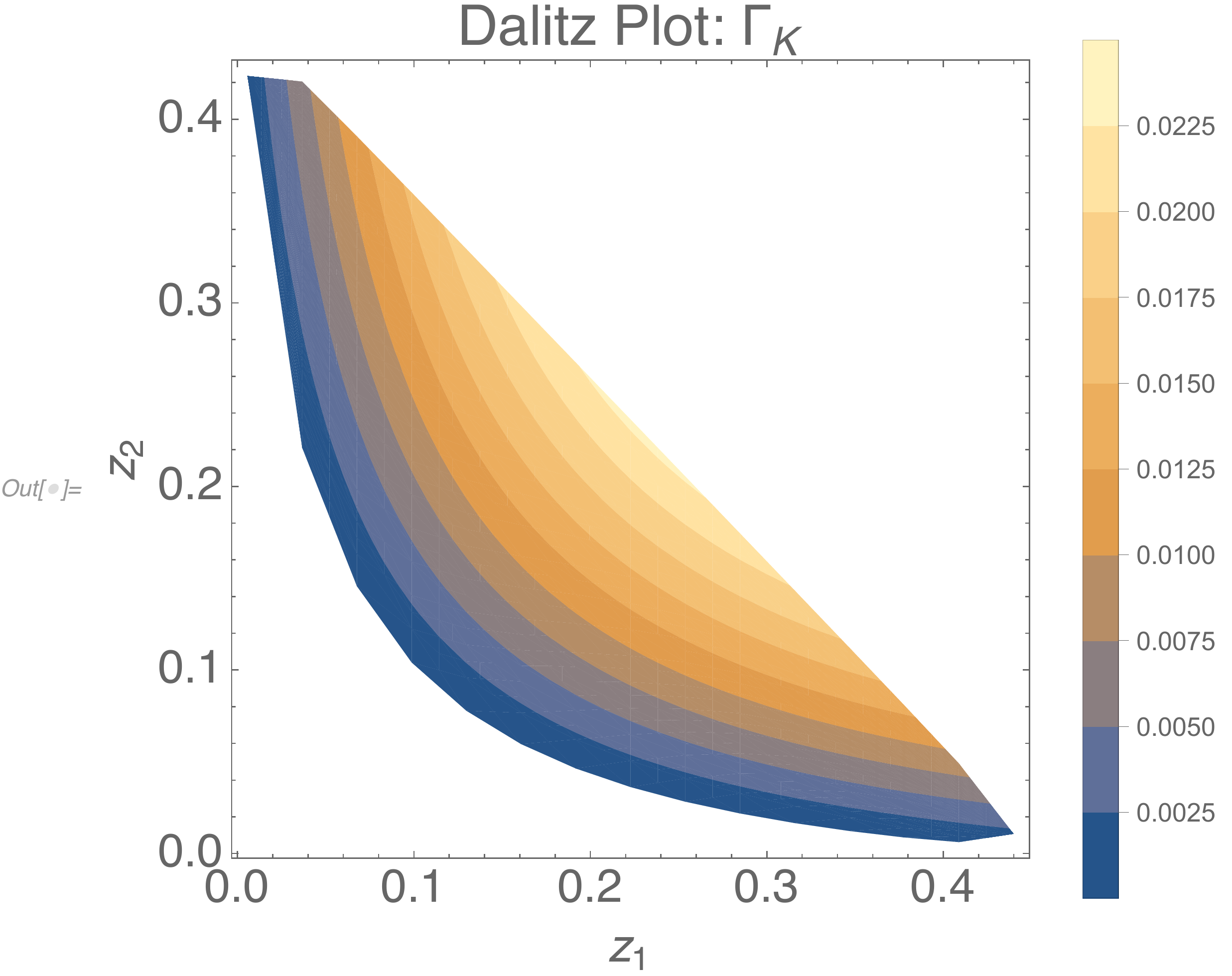}
\includegraphics[width=2.2in]{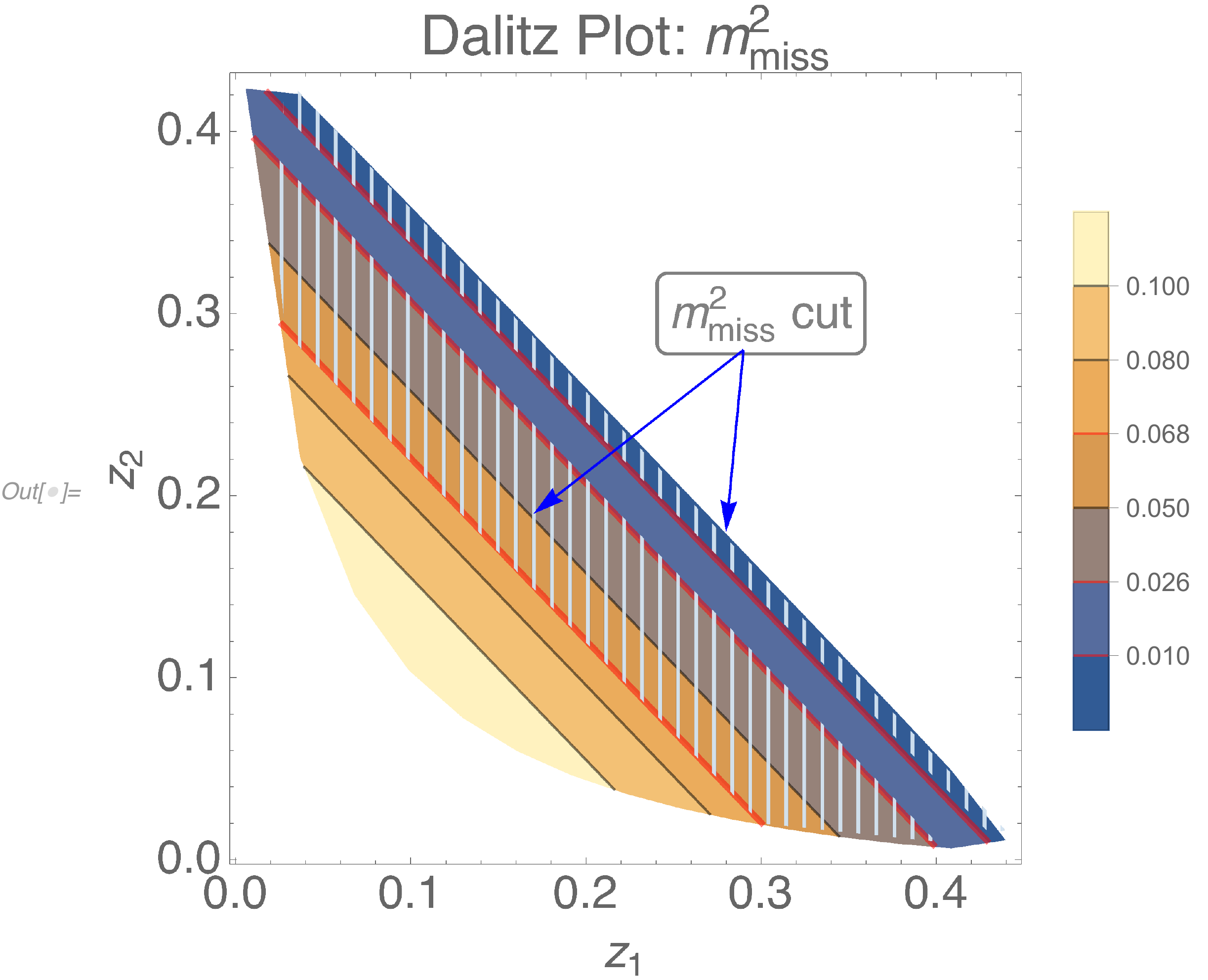}
\includegraphics[width=2.2in]{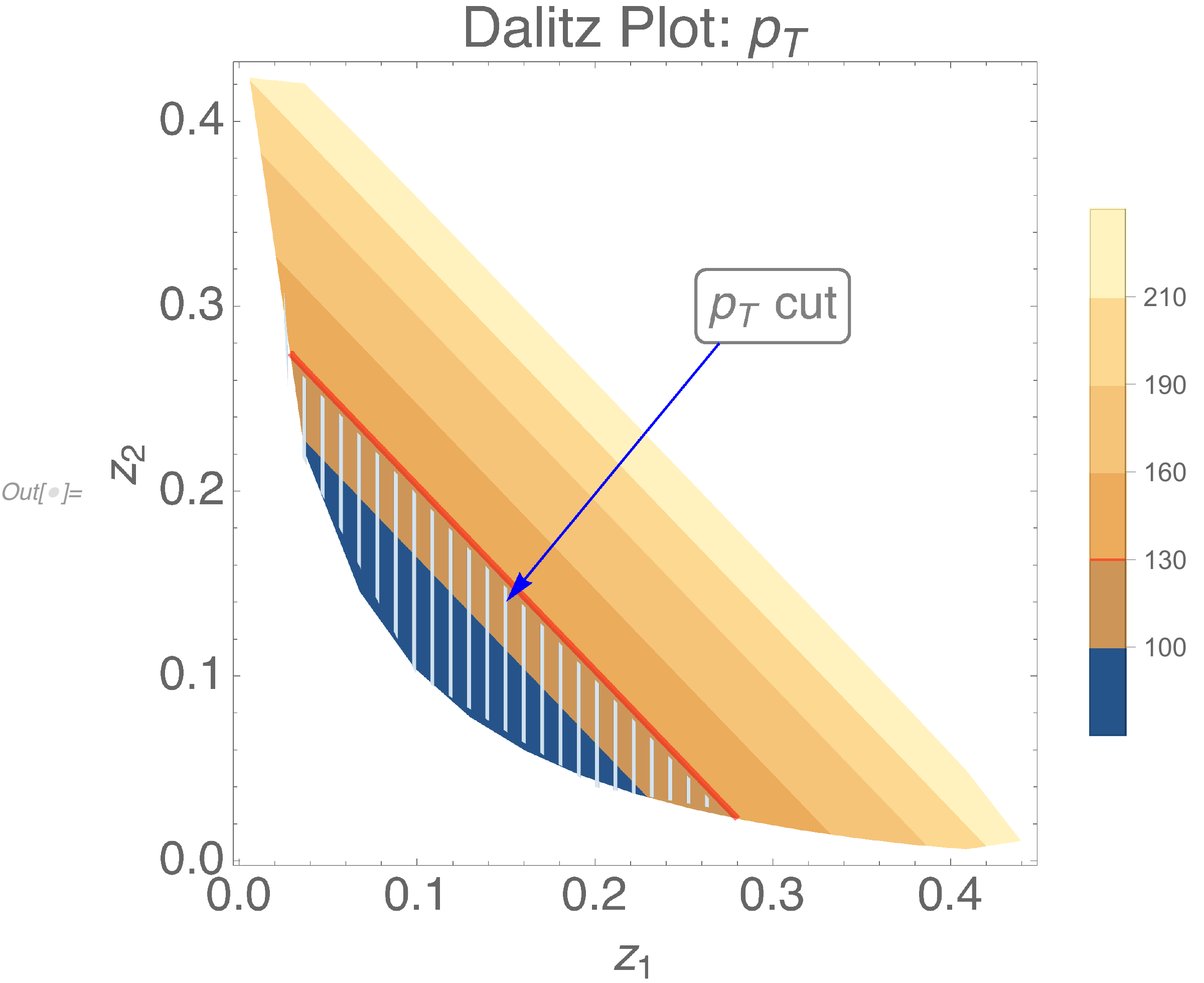}
\caption{\small 
\label{dalitz} Dalitz plots for the width $\Gamma (K^+ \rightarrow \pi^+ Q \bar Q)$ (for $m_Q=$ 10 MeV and $\alpha_D= 0.0007$), the squared missing mass $m^2_{\rm \tiny miss}$ and the transverse momentum $p_T$. Comparing the first plot on the left with those on the right, one can see how the experimental cuts for $m^2_{\rm \tiny miss}$ (NA62) remove a region (hatched between the two red contour lines and between the single red line and the upper border) where the width is at its largest while those in $p_T$ (KOTO) (the  hatched region below the red line) are less crucial.  See text for more details on the kinematic cuts.}
\end{center}
\end{figure*}

Let us now relax the strict constraints in \eq{quirky} and at the same time take into account the actual cuts implemented by the experiments in selecting the signal events.

The NA62 experiment enforces a selection  on the square of the missing mass 
\be
m^2_{\rm \tiny miss} = -m_\pi^2 + m_K^2 (1 - 2 z_1 - 2 z_2)
\ee 
and the momentum of the pion. These cuts aim to reduce the background from $K^+\rightarrow 3 \pi$ as well as to $2 \pi$. Accordingly, in order to compare the dark-sector model with experiments, we only include events within the two regions~\cite{NA62} 
\be
0.026 < m^2_{\rm \tiny miss} < 0.068\; \mbox{GeV}^2 \label{m2cut-a}
\ee
 and
 \be
 0 < m^2_{\rm \tiny miss} < 0.01\; \mbox{GeV}^2\, .\label{m2cut-b}
\ee
The momentum of the pion is taken to be between 15  and 35 GeV. 

The KOTO experiment  excludes events with a cut  on the transverse momentum 
\be 
p_T=m_K\sqrt{(r_{\pi}^2+z_1+z_2)^2-r_{\pi}^2}  \, .
\ee
 The actual cut is in part a function of the distance of the pion decay vertex~\cite{Ahn:2018mvc}; we  approximate it to a rectangular region  as
\be
130 \; \mbox{MeV}<  p_T  < 250 \; \mbox{MeV}  \label{pTcut}
\ee
and assume that the pion decays within the distance included in the experiment.


\subsection{Events selection and GN bound}

The actual number of events seen by both NA62 and KOTO is related to the BR by  the acceptances of the relative decay and the efficiency in the detection of the events.
We  look at the effect on the GN bound of   enforcing the kinematical cuts   used by the NA62 experiment on the number of events  in the case of the decay into dark-sector fermions.\footnote{A somewhat similar argument was discussed in the case of the two-body decay $K^+\to \pi^+ X^0$ in~\cite{Fuyuto:2014cya}.} 
This estimate provides only a partial inclusion of the actual differences between the dark sector  and the SM decay because the losses in the acceptance of NA62 include---on top of  the kinematical cuts in $\pi^+$ and $\Delta m_{\rm \tiny miss}^2$---also the effect of the detector geometry and particle identification and  association in the
fiducial volume. Whereas a complete analysis would require the MonteCarlo simulation of the entire experimental setup, we
only include  the  change in the  acceptance in going from the SM decay into neutrinos  to the decay into the dark-sector fermions with respect to  the kinematical cuts. This change  is a conservative estimate of the actual effect because we assume that  the efficiency of triggers and tracking as well as the overall geometric acceptance are unchanged. The change in acceptance thus included  suffices in showing that the GN bound is weakened by the experimental cuts implemented by NA62  when applied to the dark-sector decay. 

This is best understood by looking at the Dalitz plots for the decays. In Fig.~\ref{dalitz} we show the Dalitz plot for the width  $\Gamma (K^+ \to \pi^+ Q \bar Q)$ (that for  $\BR(K_L\to \pi^0 Q \bar Q )$ is the same) and compare it  with those for the kinematical variables used in the experimental cuts: $m^2_{\rm \tiny miss}$  and $p_T$. 

Because of the massless intermediate state through which the decay takes place,  the width  takes its largest values  in the  region of the Dalitz plot where $z_1$ and $z_2$ are more or less equal and in the middle of their range  (lighter color in  Fig.~\ref{dalitz}). 

Comparing the first plot  in Fig.~\ref{dalitz} with those on the right, one can see how the cuts (in red) for $m^2_{\rm \tiny miss}$ remove a region where the width is at its largest while those in $p_T$ do not.  The GN bound is not respected because of this asymmetric effect in the selection of the events after imposing  the cuts in the kinematical variables.  
 
 This feature holds true not only for the model of the dark sector we considered but  for all the models where the decay into invisible states takes place through  a light or massless intermediate state.
 Notice that if the decay were to proceed through a contact interaction---as it does in the SM---the width would be largest in the opposite range (darkest color in  Fig.~\ref{dalitz}) and the experimental cuts more symmetrical and less crucial.

\subsection{The decays in the presence of  the experimental cuts}

 \begin{figure}[h!]
\begin{center}
\includegraphics[width=3.2in]{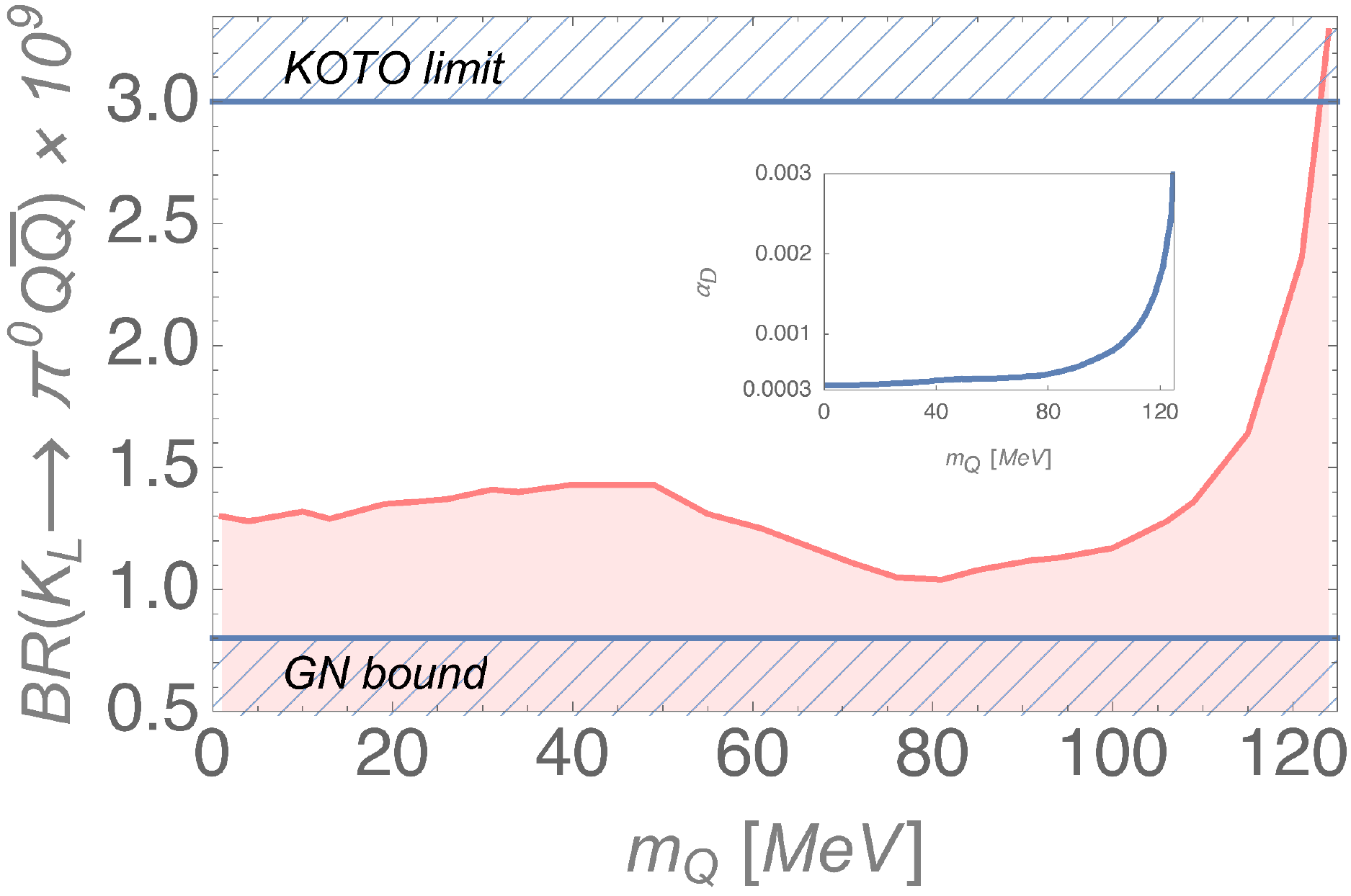}
\caption{\small Range of values for $\BR(K_L\rightarrow \pi^0 Q \bar Q )$ as function of $m_Q$. The dark gauge coupling $\alpha_D$ has been taken so as to satisfy the NA62 bound in \eq{NA62} for each value of $m_Q$. The $\BR(K_L\rightarrow \pi^0  Q \bar Q) < 3.0 \times 10^{-9}$ (KOTO limit) because of the limit in \eq{KOTOold}. Also indicated is the GN bound corresponding to \eq{NA62}. The hatched areas are excluded. The inset depicts  the values of $\alpha_{\D}$ as a function of $m_Q$ as obtained by the procedure outlined in the text.
\label{br0} }
\end{center}
\end{figure}

 In order to satisfy the  bound in \eq{NA62} for $\BR(K^+\rightarrow \pi^+ Q \bar Q )$ for values of $m_Q$ outside the range in \eq{quirky}, in which this BR is zero, we must take smaller values of $\alpha_{\D}$ with respect to the range considered in the previous section. The procedure to implement these constraints is the following.
 
For $m_{K_L}=497.611$, $m_{K^+}=493.677$, $m_{\pi^0}=134.977$ MeV and $m_{\pi^+}=139.57$~\cite{PDG}, while  again assuming as before maximal CP violation ($\sin \delta_{\mbox{\tiny CP}}=1$) and enforcing the limit in \eq{limit} from Kaon mixing, we can obtain an upper bound on $\alpha_{\D}$ by requiring that the number of events generated by the  $\BR(K^+\rightarrow \pi^+ Q \bar Q )$ satisfies the NA62 experimental bound in \eq{NA62}. This limit is computed after enforcing the experimental cuts in \eq{m2cut-a} and \eq{m2cut-b}. The value of $\alpha_{\D}$ thus  found can then be inserted, together with the corresponding value for $m_Q$, to obtain an upper bound on $\BR(K_L\rightarrow \pi^0 Q \bar Q )$. 

 Fig.~\ref{br0} shows the result of this procedure. The $\BR(K_L\rightarrow \pi^0 Q \bar Q )$ is a function of $m_Q$ and the  value of $\alpha_{\D}$  obtained by implementing the constraint in \eq{NA62} on the $\BR(K^+\rightarrow \pi^+ Q \bar Q )$.  The $\alpha_{\D}$  coupling varies within the range $0.0003 < \alpha_{\D} < 0.003$ as indicated in the inlet of Fig.~\ref{br0}; the maximum allowed value of $\alpha_{\D}$ grows quadratically as the mass $m_Q$ comes closer to the kinematical threshold. The red curve is the upper bound of the BR. The area below (indicated by the lighter red region)  covers the entire KOTO SES region (as depicted in Fig.~\ref{data})  as $m_Q$ varies between zero and  120 MeV. Larger values of $m_Q$ give a BR too large and already excluded by KOTO.

 \section{Conclusions}
 
 The recently announced new limit on the Kaon decay $K^+\rightarrow \pi^+ \nu \bar \nu$~\cite{NA62} implies that very little room is left in this channel for new physics. If the GN bound is applied, the decay $K_L\rightarrow \pi^0  \nu \bar \nu$ is constrained to be lower than most of the current {\sc KOTO} sensibility~\cite{KOTO}. The potential  tension between events  to be found  by the  {\sc KOTO} collaboration and the GN bound  can be resolved in a model of the dark sector with light dark fermions $Q$ behaving as neutrinos in the detector, via the decay channel $K_L\rightarrow \pi^0  Q \bar Q$.      
 A BR$(K_L\rightarrow \pi^0  Q \bar Q)$ above the SM prediction and  in the region currently probed by KOTO can be attained  if we take into account the asymmetric effect of the  selection of events  by the different cuts on the kinematical variables enforced by the NA62 and KOTO experiments.

\begin{acknowledgements}
       {\small
We thanks Gaia Lanfranchi for discussions on the NA62 experiments. MF is affiliated to the Physics Department of the University of Trieste and the Scuola Internazionale Superiore di Studi Avanzati---the support of which is gratefully acknowledged. 
MF and EG are affiliated to the Institute for Fundamental Physics of the Universe, Trieste, Italy.}
\end{acknowledgements}



\end{document}